\documentclass[12pt,a4paper]{article}

\usepackage{epsfig,graphicx}
\usepackage{amsmath,amsfonts,amssymb}
\usepackage{citesort}

\newcommand{\unit}{\leavevmode\hbox{\small1\kern-3.6pt\normalsize1}}

\def\lsim{\raise0.3ex\hbox{$\;<$\kern-0.75em\raise-1.1ex\hbox{$\sim\;$}}}
\def\gsim{\raise0.3ex\hbox{$\;>$\kern-0.75em\raise-1.1ex\hbox{$\sim\;$}}}

\begin{document}

\thispagestyle{empty}
\begin{flushright}
FTUAM 05/12\\
IFT-UAM/CSIC-05-33\\
KAIST-TH 2005/14\\
\vspace*{5mm}{August 27, 2005}
\end{flushright}

\begin{center}
{\Large \textbf{ 
Proposal for a 
Supersymmetric Standard Model
} 
}

\vspace{0.5cm}
\hspace*{-1mm}
D.E.~L\'opez-Fogliani$^{a,b}$ 
and C.~Mu\~noz$^{a,b,c}$\\[0.2cm]
{$^{a}$\textit{Departamento de F\'{\i}sica Te\'{o}rica C-XI, Universidad Aut\'{o}noma de Madrid,
Cantoblanco, E-28049 Madrid, Spain}}\\[0pt]
{$^{b}$\textit{Instituto de F\'{\i}sica Te\'{o}rica C-XVI, Universidad Aut\'{o}noma de Madrid,\\
Cantoblanco, E-28049 Madrid, Spain}}\\[0pt]
{$^{c}$\textit{Department of Physics, Korea Advanced Institute of
    Science and Technology,\\
Daejeon 305-701, Korea}}\\[0pt]

\begin{abstract}
The fact that neutrinos are massive suggests
that the minimal supersymmetric standard model 
(MSSM)
might
be extended in order to include three
gauge-singlet neutrino superfields with Yukawa couplings of the type
$H_2\,  L \,  \nu^c$.
We propose to use these superfields to solve the $\mu$ problem of the MSSM
without having to introduce an extra singlet superfield as in the 
case of the next-to-MSSM (NMSSM).
In particular, terms of the type $\nu^c H_1H_2$ in the superpotential 
may carry out this
task spontaneously through sneutrino vacuum expectation values.
In addition, terms of the type  $(\nu^c)^3$ avoid the presence of axions
and generate effective Majorana masses for neutrinos at the
electroweak scale.
On the other hand, these terms break lepton number and $R$-parity 
explicitly
implying that
the phenomenology of this model is very different from the one 
of the MSSM or NMSSM.
For example, the usual neutralinos are now mixed with the neutrinos.
For Dirac masses of the latter
of order $10^{-4}$ GeV, eigenvalues reproducing 
the correct scale of neutrino
masses
are obtained.
\end{abstract}
\end{center}

\vspace*{15mm}\hspace*{3mm}
\newpage


Neutrino experiments have confirmed during the last years 
that neutrinos are massive \cite{experiments}.
As a consequence, all theoretical models must be modified in order to
reproduce this result.
In particular, it is natural in the context of the Minimal Supersymmetric Standard
Model (MSSM) \cite{mssm} to supplement the ordinary  neutrino
superfields, $\hat \nu_i$, 
$i=1,2,3$, 
contained in the $SU(2)_L$-doublet, $\hat L_i$, with
gauge-singlet neutrino superfields, $\hat \nu^c_i$.
Once experiments induce us to introduce these new superfields, and given the
fact
that sneutrinos are allowed to get vacuum expectation values (VEVs),
we may wonder why not to use terms
of the type $\hat \nu^c \hat H_1\hat H_2$ 
to produce an effective  $\mu$ term.
This would allow us to solve the naturalness problem of the
MSSM, the so-called $\mu$ problem \cite{mupb}, 
without having to introduce an extra singlet superfield
as in case of the Next-to-Minimal Supersymmetric Standard 
Model (NMSSM) \cite{nmssm}.
It is true that in the model with
Bilinear $R$-parity Violation (BRpV) \cite{Valle},
the bilinear terms
$\hat H_2 \hat L_i$ induce neutrino masses 
through the mixing with
the neutralinos (actually only one mass at tree level and the other two at one loop)
without using 
the superfields $\hat \nu^c_i$, however
the $\mu$ problem is augmented
with the three new bilinear terms.

Thus the aim of this paper is to analyse the ``$\mu$ from $\nu$''
Supersymmetric Standard Model ($\mu$$\nu$SSM)  arising from
this proposal: natural particle content without $\mu$ problem.



In addition to the MSSM Yukawa couplings for quarks and charged leptons, the
$\mu$$\nu$SSM superpotential contains Yukawa couplings for neutrinos, and  
two additional type of terms involving the Higgs doublet
superfields, $\hat H_1$ and $\hat H_2$, and the three
neutrino superfields, $\hat \nu^c_i$,
%
\begin{align}\label{superpotential}
W = &
\ \epsilon_{ab} \left(
Y_u^{ij} \, \hat H_2^b\, \hat Q^a_i \, \hat u_j^c +
Y_d^{ij} \, \hat H_1^a\, \hat Q^b_i \, \hat d_j^c +
Y_e^{ij} \, \hat H_1^a\, \hat L^b_i \, \hat e_j^c +
Y_\nu^{ij} \, \hat H_2^b\, \hat L^a_i \, \hat \nu^c_j 
\right)
\nonumber\\
& 
-\epsilon_{ab} \lambda^{i} \, \hat \nu^c_i\,\hat H_1^a \hat H_2^b
+
\frac{1}{3}
\kappa^{ijk} 
\hat \nu^c_i\hat \nu^c_j\hat \nu^c_k\,,
\end{align}
where we take $\hat H_1^T=(\hat H_1^0, \hat H_1^-)$, $\hat H_2^T=(\hat
H_2^+, \hat H_2^0)$, 
$\hat Q_i^T=(\hat u_i, \hat d_i)$, 
$\hat L_i^T=(\hat \nu_i, \hat e_i)$, $a,b$ are
$SU(2)$ indices, and $\epsilon_{12}=1$.
In this model, the usual MSSM bilinear
$\mu$ term is absent from the superpotential, and only dimensionless trilinear
couplings are present in $W$.
For this to happen we can invoke a $Z_3$ symmetry 
as is usually done in the NMSSM. On the other hand,
let us recall that this is actually what happens 
in the low energy limit of string constructions: 
only trilinear couplings are present in the superpotential.
Since string theory seems to be relevant for the unification of
interactions, including gravity, this argument in favour of the
absence of a bare $\mu$ term in the superpotential is robust.
When the scalar components of the superfields $\hat\nu^c_i$,
denoted by $\tilde\nu^c_i$, acquire
VEVs of order the electroweak scale, 
an effective interaction $\mu \hat H_1 \hat H_2$ is generated
through the fifth term in  (\ref{superpotential}), with  
$\mu\equiv
\lambda^i \langle \tilde \nu^c_i \rangle$.
The last type of terms in (\ref{superpotential}) is allowed by all symmetries,
and avoids the presence of an unacceptable axion associated to a global $U(1)$
symmetry.
In addition, it generates effective Majorana masses for neutrinos at
the
electroweak scale.
These two type of terms replace the two NMSSM terms
$\hat S\hat H_1 \hat H_2$, 
$\hat S\hat S \hat S$, with $\hat S$ an extra singlet superfield. 

It is worth noticing that these terms break explicitly
lepton number, and therefore, after spontaneous symmetry breaking,
a massless Goldstone boson (Majoron)
does not appear. On the other hand,
$R$-parity (+1 for particles and -1 for superpartners)
is also explicitly broken and this means that the 
phenomenology of the $\mu$$\nu$SSM is going to be very different from the one 
of the MSSM. Needless to mention, the 
lightest $R$-odd particle is not stable.
Obviosly,
the neutralino is no longer a candidate for dark matter.
Nevertheless, other candidates can be found in the literature,
such as the
gravitino \cite{yamaguchi}, 
the well-known axion, 
and many other (exotic) particles \cite{reviewmio}. 
It is also interesting to realise that the Yukawa couplings producing 
Dirac
masses for neutrinos, the fourth term in (\ref{superpotential}),
generate through the VEVs of $\tilde\nu^c_i$, 
three effective bilinear terms
$\hat H_2 \hat L_i$.
As mentioned above these
characterize the BRpV.

Let us finally remark that
the superpotential (\ref{superpotential}) has a $Z_3$ symmetry, just like 
the NMSSM.
Therefore, one expects to have also a 
cosmological domain wall problem \cite{wall,wall2} in this model. 
Nevertheless, 
the usual solutions to this problem \cite{nowall} will also work in this 
case: non-renormalizable operators \cite{wall} in the superpotential can break 
explicitly the dangerous $Z_3$ symmetry, lifting the degeneracy of the 
three original vacua, and this can be done without introducing hierarchy 
problems. In addition, these operators can be chosen small enough as 
not to alter the low-energy phenomenology.


Working in the framework of gravity mediated supersymmetry breaking,
we will discuss now in more detail the phenomenology of the $\mu$$\nu$SSM.
Let us write first the soft terms appearing in the Lagrangian
 $\mathcal{L}_{\text{soft}}$, which in our conventions is
given by
%
%
\begin{eqnarray}
-\mathcal{L}_{\text{soft}} & =&
 (m_{\tilde{Q}}^2)^{ij} \, \tilde{Q^a_i}^* \, \tilde{Q^a_j}
+(m_{\tilde u^c}^{2})^{ij} 
\, \tilde{u^c_i}^* \, \tilde u^c_j
+(m_{\tilde d^c}^2)^{ij} \, \tilde{d^c_i}^* \, \tilde d^c_j
+(m_{\tilde{L}}^2)^{ij} \, \tilde{L^a_i}^* \, \tilde{L^a_j}
+(m_{\tilde e^c}^2)^{ij} \, \tilde{e^c_i}^* \, \tilde e^c_j
\nonumber \\
&+ &
m_{H_1}^2 \,{H^a_1}^*\,H^a_1 + m_{H_2}^2 \,{H^a_2}^* H^a_2 +
(m_{\tilde\nu^c}^2)^{ij} \,\tilde{{\nu}^c_i}^* \tilde\nu^c_j 
\nonumber \\
&+&
\epsilon_{ab} \left[
(A_uY_u)^{ij} \, H_2^b\, \tilde Q^a_i \, \tilde u_j^c +
(A_dY_d)^{ij} \, H_1^a\, \tilde Q^b_i \, \tilde d_j^c +
(A_eY_e)^{ij} \, H_1^a\, \tilde L^b_i \, \tilde e_j^c 
\right.
\nonumber \\
&+&
\left.
(A_{\nu}Y_{\nu})^{ij} \, H_2^b\, \tilde L^a_i \, \tilde \nu^c_j 
+ \text{H.c.}
\right] 
\nonumber \\
&+&
\left[-\epsilon_{ab} (A_{\lambda}\lambda)^{i} \, \tilde \nu^c_i\, H_1^a  H_2^b
+
\frac{1}{3}
(A_{\kappa}\kappa)^{ijk} 
\tilde \nu^c_i \tilde \nu^c_j \tilde \nu^c_k\
+ \text{H.c.} \right]
\nonumber \\
&-&  \frac{1}{2}\, \left(M_3\, \tilde\lambda_3\, \tilde\lambda_3+M_2\,
  \tilde\lambda_2\, \tilde
\lambda_2
+M_1\, \tilde\lambda_1 \, \tilde\lambda_1 + \text{H.c.} \right) \,.
\label{2:Vsoft}
\end{eqnarray}
%
In addition to terms from $\mathcal{L}_{\text{soft}}$, the 
tree-level scalar potential receives the usual $D$ and $F$ term
contributions.
Once the electroweak symmetry is spontaneously broken, the neutral
scalars develop
in general
the following VEVs:
\begin{equation}\label{2:vevs}
\langle H_1^0 \rangle = v_1 \, , \quad
\langle H_2^0 \rangle = v_2 \, , \quad
\langle \tilde \nu_i \rangle = \nu_i \, , \quad
\langle \tilde \nu_i^c \rangle = \nu_i^c \,.
\end{equation}
%
In what follows it will be enough for our purposes to neglect mixing
between generations in (\ref{superpotential}) and (\ref{2:Vsoft}), and
to assume that only one generation of
sneutrinos gets VEVs,
$\nu$, $\nu^c$. The extension of the analysis to all generations
is straightforward, and the conclusions are similar.
We
then obtain for the tree-level neutral scalar potential:
\begin{eqnarray}\label{2:pot}
\langle V_{\mathrm{neutral}}
\rangle 
&=&
\frac{g_1^2+g_2^2}{8} \left( |\nu|^2 + |v_1|^2 - |v_2|^2 \right)^2
\nonumber \\
&+&
|\lambda|^2 \left(
|\nu^c|^2 |v_1|^2 + |\nu^c|^2 |v_2|^2 + |v_1|^2 |v_2|^2\right) +
|\kappa|^2 |\nu^c|^4 \nonumber \\
& + & 
|Y_{\nu}|^2 \left(
|\nu^c|^2 |v_2|^2 + |\nu^c|^2 |\nu|^2 + |\nu|^2 |v_2|^2 \right)
\nonumber \\
& + &
m_{H_1}^2 |v_1|^2 + m_{H_2}^2 |v_2|^2 + m_{\tilde \nu^c}^2 |\nu^c|^2
+ m_{\tilde \nu}^2 |\nu|^2
\nonumber \\
& + &
\left(-\lambda \kappa^* v_1 v_2 {\nu^c}^{*2}
-\lambda Y_{\nu}^* |\nu^c|^2 v_1  \nu^*
-\lambda Y_{\nu}^* |v_2|^2 v_1  \nu^*
+ k Y_{\nu}^* v_2^* {\nu}^*  {\nu^c}^{2} \right.
\nonumber \\
&-& \left. \lambda A_\lambda \nu^c v_1 v_2 +
Y_{\nu} A_{\nu} \nu^c \nu v_2 
+ \frac{1}{3}
\kappa A_\kappa {\nu^c}^3 + \mathrm{H.c.}
\right) \,.
\end{eqnarray}
In the following, we assume for simplicity that all parameters
in the potential are real.
One can derive the four minimization conditions with respect to the VEVs
$v_1$, $v_2$, $\nu^c$, $\nu$, with the result
%
\begin{eqnarray}\label{2:minima}
\frac{1}{4}(g_1^2+g_2^2)(\nu^2+v_1^2-v_2^2)v_1 + \lambda^2 v_1\left( {\nu^c}^2
  + v_2^2\right)+ m_{H_1}^2 v_1 
-\lambda \nu^c v_2 \left(\kappa \nu^c  +A_\lambda \right) 
\nonumber \\
-\lambda Y_\nu \nu  \left({\nu^c}^2  + v_2^2 \right) 
= 0 \,,
\nonumber \\
-\frac{1}{4}(g_1^2+g_2^2)(\nu^2+v_1^2-v_2^2)v_2 + \lambda^2 v_2\left( {\nu^c}^2
  + v_1^2\right)  + m_{H_2}^2 v_2 
-\lambda \nu^c v_1 \left(\kappa \nu^c  +A_\lambda \right) 
\nonumber \\
+ Y_\nu^2 v_2 \left({\nu^c}^2  + \nu^2 \right) 
+ Y_\nu \nu  \left(-2 \lambda v_1v_2 + \kappa {\nu^c}^2  + A_\nu \nu^c \right) 
= 0 \,,
\nonumber \\
\lambda^2 \left(v_1^2 + v_2^2\right) \nu^c 
+2 \kappa^2 {\nu^c}^3 +
m_{\tilde\nu^c}^2  \nu^c
-2\lambda\kappa v_1 v_2 \nu^c 
-\lambda A_{\lambda} v_1v_2 + \kappa A_{\kappa} {\nu^c}^2
\nonumber \\
+ Y_\nu^2 \nu^c  \left(v_2^2  + \nu^2 \right) 
+ Y_\nu \nu  
\left(-2 \lambda \nu^c v_1 
+ 2\kappa v_2 {\nu^c}  + A_\nu v_2\right) 
=0 \,,
\nonumber \\
\frac{1}{4}(g_1^2+g_2^2)(\nu^2+v_1^2-v_2^2)\nu + m_{\tilde\nu}^2 \nu
\nonumber \\
+ Y_\nu^2 \nu \left( v_2^2 +  {\nu^c}^2   \right)   
+ Y_\nu 
\left(-\lambda {\nu^c}^2 v_1  -\lambda v_2^2 v_1  
+ \kappa v_2 {\nu^c}^2  + A_\nu \nu^c v_2\right) 
= 0 \,.
\end{eqnarray}
As discussed in the context of $R$-parity breaking models with extra
singlets \cite{Masiero},
the VEV of the left-handed sneutrino,
$\nu$, is in general small.
Here we can use the same argument.
Notice that in the last equation in (\ref{2:minima})
$\nu\to 0$ as $Y_{\nu}\to 0$, and since
the coupling $Y_{\nu}$ determines the Dirac mass for the
neutrinos,  $m_D \equiv Y_{\nu}v_2$, $\nu$ has to be very small.
Using this rough argument we can also get an estimate of the value,
$\nu\lsim m_D$.
This also implies that we can approximate the other three equations as
follows:
\begin{align}\label{2:minima2}
\frac{1}{2} M_Z^2 \cos 2\beta
+\lambda^2 \left( {\nu^c}^2 + v^2\sin^2\beta \right)
+
m_{H_1}^2 
-\lambda {\nu^c} \tan \beta \left(\kappa {\nu^c} +A_\lambda \right) 
= & 0
\,,
\nonumber \\
-\frac{1}{2} M_Z^2 \cos 2\beta
+\lambda^2 \left( {\nu^c}^2 + v^2\cos^2\beta \right)
+
m_{H_2}^2 
-\lambda {\nu^c} \cot \beta \left(\kappa {\nu^c} +A_\lambda \right) 
= & 0
\,,
\nonumber \\
\lambda^2 v^2+ 2\kappa^2 {\nu^c}^2
+
m_{{\tilde\nu^c}}^2   
 - \lambda \kappa v^2
\sin 2\beta -\frac{\lambda A_\lambda v^2}{2{\nu^c}} \sin 2\beta +
\kappa A_\kappa {\nu^c}\
=
& 0\,,
\end{align}
where $\tan \beta \equiv v_2/v_1$,
$2 M_W^2/g_2^2= v_1^2+v_2^2 + \nu^2\approx v_1^2+v_2^2\equiv v^2$,
and
we have neglected terms proportional to $Y_{\nu}$.
It is worth noticing that these equations are the same as the ones
defining the minimization conditions for the NMSSM, with the 
substitution ${\nu^c}\leftrightarrow s$.
Thus one can carry out the analysis of the model
similarly to the NMSSM case, where many solutions in the parameter space
$\lambda ,\kappa ,\mu (\equiv\lambda s) , \tan \beta , A_\lambda ,
A_\kappa ,
$
can be found \cite{nmssm2}.

Once we know that solutions are available in this model,
we have to discuss in some detail the important issue of mass
matrices.
Concerning this point, the breaking of $R$-parity makes the $\mu$$\nu$SSM
very different from the MSSM and the NMSSM.
In particular, neutral gauginos and Higgsinos are now mixed
with the neutrinos.
Not only the
fermionic component of $\tilde \nu^c$ mixes with the neutral Higgsinos
(similarly to the fermionic component of $S$ in the NMSSM), but
also the fermionic component of 
$\tilde \nu$ enters in the game,
giving rise to a
sixth state. 
Of course, now we have to be sure that one eigenvalue of this matrix
is very small, reproducing the experimental results about neutrino masses.
In the weak interaction basis defined by ${\Psi^0}^T
\equiv \left(\tilde B^0=-i \tilde \lambda^\prime, \tilde W_3^0=-i
  \tilde \lambda_3, \tilde
H_1^0, \tilde H_2^0, \nu^c, \nu \right)$,  the neutral fermion mass terms in the
Lagrangian are
$\mathcal{L}_{\mathrm{neutral}}^{\mathrm{mass}}
=
-\frac{1}{2} (\Psi^0)^T \mathcal{M}_{\mathrm{n}}
\Psi^0 + \mathrm{H.c.}$,
with $\mathcal{M}_{\mathrm{n}}
$ a $6 \times 6$ ($10 \times 10$ if
we include all generations of neutrinos) matrix,
{\footnotesize \begin{equation}
  \mathcal{M}_{\mathrm{n}} = \left(
    \begin{array}{cc}
M & m \\
m^T & 0 \\
\end{array} \right),
  \label{neumatrix}
\end{equation}}
where 
{\footnotesize \begin{equation}
  M
= \left(
    \begin{array}{ccccc}
      M_1 & 0 & 
-M_Z \sin \theta_W \cos \beta 
&
M_Z \sin \theta_W \sin \beta 
& 0  \\
      0 & M_2 & 
M_Z \cos \theta_W \cos \beta 
&
      -M_Z \cos \theta_W \sin \beta 
& 0 \\
      -M_Z \sin \theta_W \cos \beta &
      M_Z \cos \theta_W \cos \beta &
      0 & -\lambda \nu^c & -\lambda v_2 \\
      M_Z \sin \theta_W \sin \beta &
      -M_Z \cos \theta_W \sin \beta &
      -\lambda \nu^c &0  & -\lambda v_1 + Y_\nu \nu \\
      0 & 0 &  -\lambda v_2 & -\lambda v_1+ Y_\nu \nu & 2 \kappa \nu^c
    \end{array} \right),
  \label{neumatrix2}
\end{equation}}
is very similar to the neutralino mass matrix of the NMSSM 
(substituting  ${\nu^c}\leftrightarrow s$ and neglecting
the contributions $Y_\nu \nu$),
and 
\begin{equation}
 {m}^T
=
\left(\,\,\,\,\,\,  -\frac{g_1\nu}{\sqrt 2}\,\,\,\,\,\,   \frac{g_2\nu}{\sqrt 2}\,\,\,\,\,\,
  0\,\,\,\,\,\,   
Y_\nu\nu^c\,\,\,\,\,\,  
Y_\nu v_2\,\,\,\,\,\,     \right)\ .
  \label{neumatrix3}
\end{equation}
Matrix (\ref{neumatrix}) is a matrix of the see-saw type that will
give rise to a very light eigenvalue if
the entries of the matrix $M$ are much larger than the entries
of the matrix $m$.
This is generically the case since the entries of $M$ are of order the
electroweak scale, but for the entries of $m$, $\nu$ is small 
and $Y_\nu v_2$ is the Dirac mass for the neutrinos $m_D$ 
as discussed above
($Y_\nu\nu^c$ has the same order of magnitude of $m_D$).
We have checked numerically that correct neutrino masses can easily be
obtained.
For example, using typical electroweak-scale values in (\ref{neumatrix2}), 
and a Dirac mass of order $10^{-4}$ GeV
in (\ref{neumatrix3}),
one obtains that the lightest eigenvalue of (\ref{neumatrix}) 
is of order $10^{-2}$ eV.
Including the three generations in the analysis we can obtain different
neutrino mass hierarchies playing with the 
hierarchies in the Dirac masses.

The possibility of using a 
see-saw at the electroweak scale has not been
considered in much detail in the literature [For a recent
work see ref.~\cite{kitano}, where an extension of the NMSSM is 
considered with Majorana masses for neutrinos generated dynamically
through the VEV of the singlet $S$. R-parity may be broken in this
extension, although spontaneously], although this 
avoids the introduction of ad-hoc high energy scales. 
Of course, with a see-saw at the scale of a Grand Unified Theory (GUT)  
one can have Yukawa couplings of order one for neutrinos.
However, since we know that the Yukawa coupling of the electron has to be
of order $10^{-6}$, why the one of the neutrino should be
six orders of magnitude larger ?
As mentioned above, with the electroweak-scale see-saw 
a Yukawa coupling of order of the one of the electron is
sufficient to reproduce the neutrino mass.
Notice also that a purely Dirac mass for the neutrino 
would imply a Yukawa coupling of order $10^{-13}$,
i.e. seven orders of magnitude smaller than the one we need
with a electroweak-scale see-saw.
It is worth mentioning here that in some string constructions, where
supersymmetric standard-like models can be obtained without the necessity
of a GUT, and
Yukawa couplings can be explicitly computed, those for neutrinos
cannot be as small as $10^{-13}$, and therefore the presence of a see-saw
at the electroweak scale is helpful \cite{AbelMunoz}.
In any case, let us remark that in our model the see-saw is
dynamical and unavoidable, since the matrix of eq. (\ref{neumatrix}) 
producing such a see-saw is
always present.

It has been noted
in the literature that the sneutrino-antisneutrino mixing effect
generates a loop correction to the neutrino mass, which
depends on the
mass-splitting of the sneutrino mass eigenstates \cite{Haber}. In the
case of assuming a
large Majorana mass this correction is negligible if all parameters are
of order the supersymmetric scale.
We have checked that the same result is obtained in our model
with a see-saw at the electroweak scale, unless a fine tune of the 
parameters is forced producing a too large sneutrino mass difference.



On the other hand, the charginos mix with the charged leptons
and therefore in a basis where ${\Psi^+}^T
\equiv \left(-i \tilde \lambda^+, \tilde H_2^+ , 
e_R^+\right)$
and 
${\Psi^-}^T
\equiv \left(-i \tilde \lambda^-, \tilde H_1^- , 
e_L^-\right)$,
one obtains the matrix 
{\footnotesize \begin{equation}
\left(
    \begin{array}{ccc}
M_2 & g_2 v_2 & 0\\
g_2 v_1 & \lambda \nu^c & -Y_e \nu \\
g_2\nu  & -Y_\nu \nu^c  & Y_e v_1
\end{array} \right).
  \label{neumatrixch}
\end{equation}}
Here we can distinguish the $2 \times 2$ submatrix which
is similar to the chargino mass matrix of the NMSSM 
(substituting  ${\nu^c}\leftrightarrow s$).
Clearly, given the vanishing value of the 13 element of the
matrix (\ref{neumatrixch}), and the extremely small absolute value of 
the 23 element, 
there will always be a light eigenvalue corresponding to the 
electron mass $Y_e v_1$. 
The extension of the analysis to three generations is again
straightforward.
 
Of course, other mass matrices are also modified.
This is the case for example of the Higgs boson mass matrices.
The presence of the VEVs $\nu$, $\nu^c$, leads to mixing of the
neutral Higgses with the sneutrinos. 
Concerning the Higgs phenomenology,
since basically the $\nu^c$ plays the role
of the singlet $S$,
this will be similar to the one of the NMSSM \cite{nmssm2}. For example, 
two CP-odd Higgses are present, and 
we have checked that one of them can in principle be light.
Likewise the
charged Higgses will be mixed with the charged sleptons.
On the other hand,
when compared to the MSSM case, the structure of squark mass terms
is essentially unaffected, provided that one uses $\mu = \lambda\nu^c
$,
and neglects the contribution of the fourth term in (\ref{superpotential}).

Obviously, the phenomenology of the $\mu$$\nu$SSM is very rich and different
from other models, and therefore many more issues might have been addressed,
such as possible experimental constraints,
implications for accelerator physics, analysis of the (modified)
renormalization group equations, study of the neutrino
masses in detail, etc.
However, these are beyond the scope of this paper, and we leave this necessary
task for a future work \cite{preparation}.
Our main interest here was to introduce the characteristics of this
new model, and sketch
some important points concerning its phenomenology.

  \section*{Acknowledgements}\label{ack}
 D.E. L\'opez-Fogliani is grateful to the members of the Theoretical
 High Energy Physics Group at KAIST for their kind hospitality,
when this project was started.
 He also acknowledges the financial support of the
 Spanish DGI through a FPU grant.\
 The work of C. Mu\~noz was supported 
 in part by the Spanish DGI of the
 MEC under Proyectos Nacionales BFM2003-01266 and FPA2003-04597;
 also by the European Union under the RTN program  
 MRTN-CT-2004-503369, and under the ENTApP Network of the ILIAS
 project
 RII3-CT-2004-506222; and also by KAIST under the Visiting Professor Program.


\begin{thebibliography}{99}


\bibitem{experiments} Super-Kamiokande collaboration,
Y. Fukuda et al., 
{\it Phys. Rev. Lett.} {\bf 81} (1998) 1562 [arXiv:hep-ex/9807003];
SNO collaboration, Q.R. Ahmad et al.,
{\it Phys. Rev. Lett.} {\bf 89} (2002) 011301 [arXiv:nucl-ex/0204008];
KamLAND collaboration, K. Eguchi et al.,
{\it Phys. Rev. Lett.} {\bf 90} (2003) 021802 [arXiv:hep-ex/0212021].


\bibitem{mssm} For reviews, see
H.P. Nilles, 
{\it Phys. Rep.} {\bf 110} (1984) 1;
H.E. Haber and G.L. Kane,
{\it Phys. Rep.} {\bf 117} (1985) 75.


\bibitem{mupb} J.E. Kim and H.P. Nilles, 
{\it Phys. Lett.} {\bf B138} (1984) 150.


\bibitem{nmssm} 
P. Fayet, {\it Nucl.\ Phys.\ } {\bf B90} (1975) 104;
H.~P.~Nilles, M.~Srednicki and D.~Wyler, 
{\it Phys.\ Lett.\ } {\bf B120} (1983)
346;
%
J.~M.~Frere, D.~R.~T.~Jones and S.~Raby, 
{\it Nucl.\ Phys.\ } {\bf B222} (1983) 11;
%
J.~P.~Derendinger and C.~A.~Savoy, 
{\it Nucl.\ Phys.\ } {\bf B237} (1984) 307;
J.~R.~Ellis, J.~F.~Gunion, H.~E.~Haber, L.~Roszkowski and F.~Zwirner,
{\it Phys.\ Rev.\ } {\bf D39} (1989) 844;
%
\noindent M.~Drees, 
{\it
Int.\ J.\ Mod.\ Phys.\ } {\bf A4} (1989) 3635;
%
\noindent U.~Ellwanger, M.~Rausch de Traubenberg and C.~A.~Savoy, 
{\it Phys.\ Lett.\ }
{\bf B315} (1993) 331 [arXiv:hep-ph/9307322]; 
%
P.N. Pandita, {\it Phys.\ Lett.\ }
{\bf B318} (1993) 338,
{\it Z. Phys.}
{\bf C59} (1993) 575;
\noindent S.~F.~King and P.~L.~White, 
{\it Phys.\ Rev.\ } {\bf D52}
(1995) 4183 [arXiv:hep-ph/9505326];
\noindent U.~Ellwanger and C. Hugonie, 
{\it Eur. Phys.\ J.\ }
{\bf C13} (2000) 681 [arXiv:hep-ph/9812427].


\bibitem{Valle} See e.g., M. Hirsch and J.W.F. Valle,
{\it New J. Phys.} {\bf 6} (2004) 76 [arXiv:hep-ph/0405015],
and references therein.




\bibitem{yamaguchi} For analyses of gravitino dark matter
without R-parity, see: F. Takayama and M. Yamaguchi,
{\it Phys. Lett.} {\bf B485} (2000) 388 
[arXiv:hep-ph/0005214];
M. Hirsch, W. Porod and D. Restrepo,
{\it J. High Energy Phys.} {\bf 03} (2005) 062 [arXiv:hep-ph/0503059].



\bibitem{reviewmio} For a recent review, see:
C. Mu\~noz,
{\it Int. J. Mod. Phys.} {\bf A19} (2004) 3093 
[arXiv:hep-ph/0309346].



%
\bibitem{wall}
J.~R.~Ellis, K.~Enqvist, D.~V.~Nanopoulos, K.~A.~Olive, M.~Quiros and 
F.~Zwirner,
{\it Phys.\ Lett.} {\bf B176} (1986) 403;
B. Ray and G. Senjanovic, 
{\it Phys.\ Rev.} {\bf D49} (1994) 2729 [hep-ph/9301240].

\bibitem{wall2}
S.~A.~Abel, S.~Sarkar and P.~L.~White,
{\it  Nucl.\ Phys.} {\bf B454} (1995) 663
[hep-ph/9506359].
%

\bibitem{nowall}
S.~A.~Abel,
{\it Nucl.\ Phys.} {\bf B480} (1996) 55 [hep-ph/9609323];
C.~Panagiotakopoulos and K.~Tamvakis,
{\it Phys.\ Lett.} {\bf B446} (1999) 224 [hep-ph/9809475].
%


\bibitem{Masiero} A. Masiero and J.W.F. Valle,
{\it Phys. Lett.} {\bf B251} (1990) 273.


\bibitem{nmssm2} See e.g., 
D.G. Cerde\~no, C. Hugonie, D.~E.~L\'opez-Fogliani,
C.~Mu\~noz and A.~M.~Teixeira, 
{\it J. High Energy Phys.} {\bf 12} (2004) 048 [arXiv:hep-ph/0408102],
and references therein.



\bibitem{kitano} R. Kitano and K.Y. Oda,
{\it Phys. Rev.} {\bf D61} (2000) 113001 [arXiv:hep-ph/9911327].


\bibitem{AbelMunoz} S.A. Abel and C. Mu\~noz,
{\it J. High Energy Phys.} {\bf 02} (2003) 010 [arXiv:hep-ph/0212258].

\bibitem{Haber} Y. Grossman and H.E. Haber,
{\it Phys. Rev. Lett.} {\bf 78} (1997) 3438 [arXiv:hep-ph/9702421];
{\it Phys. Rev.} {\bf D59} (1999) 093008 [arXiv:hep-ph/9810536];
arXiv:hep-ph/9906310.


\bibitem{preparation} D.~E.~L\'opez-Fogliani,
C.~Mu\~noz and R. Ruiz de Austri, in preparation.






































































\end{thebibliography}
\end{document}